\documentclass[reprint,floatfix,twocolumn,superscriptaddress,showpacs, aps, prl,longbibliography]{revtex4-2} 

\usepackage{graphicx} 
\usepackage{natbib} 
\usepackage[usenames,dvipsnames]{color} 
\usepackage{amsmath} 
\usepackage[urlcolor=blue, hyperindex, colorlinks, bookmarks=true,linkcolor=black,citecolor=black]{hyperref} 
\usepackage{dcolumn}
\usepackage{amssymb} 
\usepackage{soul} 
\usepackage{ifthen} 
\usepackage{bbm}
\usepackage{comment}
\usepackage[ampersand]{easylist}
\usepackage{todonotes}
\usepackage{upgreek}
\usepackage{float}
\usepackage{physics}

\def\l{\left}
\def\r{\right}

\def\be#1\ee{\begin{equation}#1\end{equation}}
\def\ba#1\ea{\begin{align}#1\end{align}}
\def\bg#1\eg{\begin{gather}#1\end{gather}}
\def\t{\text}

\def\shownoteal{0} 
\newcommand{\nal}[1]{\ifthenelse{\shownoteal=1}{\textcolor{red}{[[#1]]}}{}}
\newcommand{\nnj}[1]{\ifthenelse{\shownoteay=1}{\textcolor{orange}{[[#1]]}}{}}
\def\showaddmat{1} 
\newcommand{\addmat}[1]{\ifthenelse{\showaddmat=1}{\textcolor{Gray}{[[#1]]}}{}}
\def\shownote{1} 
\newcommand{\note}[1]{\ifthenelse{\shownote=1}{\textcolor{Red}{[[#1]]}}{}}
\begin{document}

\title{Tunable Coupler for Mediating Interactions between a Two-Level System and a Waveguide from a Decoupled State to the Ultra-Strong Coupling Regime}

\author{N. Janzen}
\affiliation{Institute for Quantum Computing, Department of Physics
and Astronomy, and Waterloo Institute for Nanotechnology, University
of Waterloo, Waterloo, ON, Canada N2L 3G1}

\author{X. Dai}
\affiliation{Institute for Quantum Computing, Department of Physics
and Astronomy, and Waterloo Institute for Nanotechnology, University
of Waterloo, Waterloo, ON, Canada N2L 3G1}

\author{S. Ren}
\affiliation{Institute for Quantum Computing, Department of Physics
and Astronomy, and Waterloo Institute for Nanotechnology, University
of Waterloo, Waterloo, ON, Canada N2L 3G1}

\author{J. Shi}
\affiliation{Institute for Quantum Computing, Department of Physics
and Astronomy, and Waterloo Institute for Nanotechnology, University
of Waterloo, Waterloo, ON, Canada N2L 3G1}

\author{A. Lupascu}
\affiliation{Institute for Quantum Computing, Department of Physics
and Astronomy, and Waterloo Institute for Nanotechnology, University
of Waterloo, Waterloo, ON, Canada N2L 3G1}

\date{ \today}

\begin{abstract}
Two-level systems (TLS) coupled to waveguides are a fundamental paradigm for light-matter interactions and quantum networks. We introduce and experimentally demonstrate a method to tune the interaction between a TLS, implemented as a flux qubit, and a transmission line waveguide from a decoupled state to a coupling strength that is a significant fraction of the TLS transition frequency, near the ultra-strong coupling regime. The coupling, controlled via magnetic flux, is described by a normalized coupling strength $\alpha$ that is measured to range between $6.2\times10^{-5}$ and $2.19\times10^{-2}$, with larger attainable maximum values predicted by a circuit model of the device. This system enables future investigations in the dynamics of the spin-boson model, microwave photonics, and relativistic quantum information.

\end{abstract}
\maketitle

\section{Introduction}\label{sec:intro}

The investigation of light-matter interactions across a wide range of coupling strengths implemented through specially designed electromagnetic environments is a growing area of research~\cite{leggett_dynamics_1987,haroche_cavity_1989,yoshihara_superconducting_2017,blais_circuit_2021}. At substantially large coupling strength, the interaction enters the ultra-strong coupling (USC) regime~\cite{kockum_ultrastrong_2019, forn-diaz_ultrastrong_2019} which enables new research avenues in several fields of study including the spin-boson model~\cite{forn-diaz_ultrastrong_2017}, quantum computing~\cite{romero_ultrafast_2012,wang_ultrafast_2017}, and relativistic quantum information~\cite{valentini_non-local_1991,reznik_violating_2005,sabin_dynamics_2010,sabin_extracting_2012,salton_acceleration-assisted_2015,pozas-kerstjens_harvesting_2015}. In superconducting devices, the USC regime was previously demonstrated for a two-level system (TLS) coupled to either single cavity modes~\cite{yoshihara_superconducting_2017}, or to a continuum of modes~\cite{forn-diaz_ultrastrong_2017}. 

Past implementations of strong coupling between a TLS and a continuum of modes in a waveguide were limited by the coupling range being either fixed or bound to a relatively narrow range. In this article, we propose and demonstrate a device capable of mediating the light-matter coupling between a two-level system and a waveguide, achieving a large range of interaction strengths from a decoupled state to coupling that approaches the USC regime. The normalized coupling strength $\alpha$ is measured to range from $6.2\times10^{-5}$ to $2.19\times10^{-2}$, which enables exploring the dynamics of the interactions over a span of several qualitatively distinct regimes, including near the USC regime where the relaxation rate becomes comparable to the splitting frequency and the rotating wave approximation (RWA) breaks down \cite{jaynes_comparison_1963,peropadre_nonequilibrium_2013}. The design of the coupler is compatible with fast-switching of the interaction, enabling new experiments in spin-boson model physics~\cite{leggett_dynamics_1987} and relativistic quantum information~\cite{sabin_dynamics_2010,sabin_extracting_2012}.

The article is organized as follows. In Sec.~\ref{sec:model}, we present a detailed discussion of our system, including a circuit model derivation of the complete system, formed of a qubit, the coupler, and a transmission line, and a qualitative model for the coupler based on magnetic susceptibility. In Sec.~\ref{sec:CalandResults}, we present experimental results, including the microfabrication of the device, the validation of the circuit model, and the measurement of the coupling performed using scattering experiments. In Sec.~\ref{sec:discussion}, we compare the coupling strength and decoherence rate with the predicted values based on the circuit model and typical noise channels. Finally, in Sec.~\ref{sec:summary}, we summarize the results and discuss several potential research applications for this device.
 
\section{Device Modeling}\label{sec:model}

\subsection{Susceptibility Description of Tunable Coupler} 

\begin{figure}
    \centering
    \includegraphics{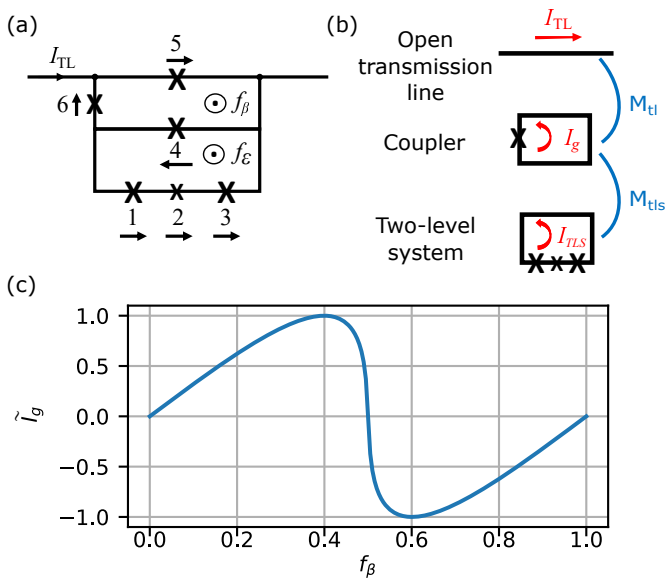}
    \caption[Circuit Diagram and Susceptibility Model]{(a) Circuit diagram of the system, showing the coupler (top) and qubit (bottom) loop, externally biased with normalized magnetic fluxes $f_\beta$ and $f_\epsilon$ respectively. The crosses represent Josephson junctions. (b) A simplified model showing the coupling between the two-level system, with persistent current $I_\t{TLS}$, and the transmission line. Refer to the text for further details. (c) Normalized ground state current $\Tilde{I_g}$ of the coupler loop versus the applied flux $f_\beta$ demonstrating the tunable susceptibility $1/L_\beta=\frac{1}{\Phi_0}\partial I_g/\partial f_\beta$ of the coupler.}
    \label{fig:Circuit}
\end{figure}
 
The system we investigate is formed of a TLS, implemented by a superconducting persistent current qubit (PCQ)~\cite{mooij_josephson_1999,orlando_superconducting_1999}, coupled to a waveguide, realized using a superconducting coplanar waveguide transmission line (TL). The device is shown schematically in Fig.~\ref{fig:Circuit}(a). It is formed of two superconducting loops with six Josephson junctions. The qubit loop, biased with normalized magnetic flux $f_\epsilon=\frac{\Phi_\epsilon}{\Phi_0}$ where $\Phi_0$ is the magnetic flux quantum, implements the TLS. The coupling loop, biased with the normalized flux $f_\beta$, effectively acts as a tunable inductor between the qubit loop and the transmission line carrying current $I_\text{TL}$. Qualitatively, the interaction between the qubit loop and the TL is analogous to previously explored devices used to tune interactions between two flux qubits \cite{brink_mediated_2005,van_der_ploeg_controllable_2007,weber_coherent_2017} and can be intuitively understood in terms of the simplified model shown in Fig.~\ref{fig:Circuit}(b). The tunable inductance between the TLS and the TL is given by $M_\text{eff} = M_\text{tls}M_\text{tl}/L_\beta$, where $M_\text{tls}$ ($M_\text{tl}$) is the mutual inductance between the coupling loop and TLS (TL) and $1/L_\beta$ is the susceptibility of the coupling loop. The mutual inductances $M_\text{tls}$ and $M_\text{tl}$ are predominantly supplied by the inductance of the shared junctions 4 and 5 respectively. The susceptibility is $1/L_\beta=\frac{1}{\Phi_0}\partial I_g/\partial f_\beta$, with $I_g$ the coupler loop ground state current, which can be tuned by changing $f_\beta$ as shown in Fig.~\ref{fig:Circuit}(c). We emphasize that this system enables a large range of coupling that goes from effectively zero when $\partial I_g/\partial f_\beta=0$ at $f_\beta\approx0.4~\t{or}~0.6$, to a maximum at $f_\beta=0.5$. This qualitative model is a simplified description for the coupler; we will introduce the full model used in this work in Sec.~\ref{sec:CircuitModel}. We note that while previous work \cite{peropadre_switchable_2010} discussed switchable coupling circuits, our design is relatively simpler as it requires only one flux control to tune the coupling.

\subsection{Light-Matter Interaction}\label{sec:SB}

The TLS, which is identified based on the lowest two energies of the double loop device, has the effective Hamiltonian
\begin{equation}
    H = -\frac{\hbar\Delta}{2}\tau_x - \frac{\hbar\epsilon}{2}\tau_z,
\end{equation}
where $\Delta$ is the minimum gap, $\tau_x$ and $\tau_z$ are the Pauli matrices, and $\epsilon = 2I_\t{TLS}\Phi_0(f_\epsilon-f_{\epsilon,\text{sym}})/\hbar$ with $I_\t{TLS}$ the persistent current in this loop. This Hamiltonian is expressed in the flux basis, formed of states with opposing currents in the qubit loop. The normalized flux $f_{\epsilon,\text{sym}}$ is the setting for which the potential energy defining the TLS has a symmetric double well shape. For a standard PCQ, formed of a single loop with three junctions, $f_{\epsilon,\text{sym}} = 0.5$. Our device has a symmetry point that is in general not at 0.5 biasing, due to the fact that the coupling loop effectively induces an additional flux component in the qubit loop. The qubit transition frequency close to the symmetry point is given by
\begin{equation} \label{eq:omega10}
            \omega_{10} = \sqrt{\Delta^2+\epsilon^2}.
\end{equation}

The Hamiltonian of the interacting qubit and transmission line is $H_\text{SB}=H_{s}+H_{b}+H_\text{int}$,
with $H_{s}$, $H_{b}$, and $H_\text{int}$ the qubit, TL, and interaction Hamiltonians.
We have
\begin{align}
    H_{s} &= -\frac{\hbar\omega_{10}}{2}\sigma_z,\\
    H_b &=\sum_k \hbar\omega_k a_k^\dagger a_k,\,\text{and}\\
    H_\text{int} &=\sum_k (g_k^x\sigma_x +g_k^z\sigma_z)(a_k^\dagger+a_k)\label{eq:int}.
\end{align}
In these equations $\omega_{10}$ is the transition frequency of the TLS, $\sigma_z$ and $\sigma_x$ are the system's Pauli operators in the energy eigenbasis, $\omega_k$, $a_k^\dagger$, $a_k$ are frequency, creation, and annihilation operators of the $k^{\text{th}}$ mode of the bosonic bath, and $g_k^{x(z)}$ are the transverse (longitudinal) coupling of the TLS to mode $k$. More details of the derivation of the Hamiltonian is provided in the Appendix~\ref{sec:SBFromCircuit}.

When the qubit is at the symmetry point, the interaction with the TL is dominantly transverse and the Hamiltonian becomes the standard spin-boson model~\cite{leggett_dynamics_1987,weiss_quantum_2012}. The environment coupling is captured by the bath spectral density function $J(\omega)=\pi\alpha\omega$, where $\alpha$ is the dimensionless coupling factor to the bath. For small values of $\alpha \ll 1$ at the symmetry point, the TLS radiative relaxation rate $\Gamma_{1}$ is given by 
\begin{equation} \label{eq:alpha}
    \Gamma_{1} = \pi\alpha\Delta.
\end{equation}

\subsection{Circuit Model}\label{sec:CircuitModel}

In this section, we introduce the Hamiltonian of the coupler-qubit circuit and discuss how it is used to extract the coupling strength from the system. The phases across the junctions of the circuit shown in Fig.~\ref{fig:Circuit}(a) are constrained by the fluxoid quantization condition, given by
\begin{align} \label{eq:flux_rel}
\begin{split}
    \gamma_1+\gamma_2+\gamma_3+\gamma_4+2\pi f_\epsilon &= 0~~\text{and}\\
    \gamma_4+\gamma_5+\gamma_6 - 2\pi f_\beta &= 0,
\end{split}
\end{align}
where $\gamma_i$ is the phase across junction $i$, and $f_{\epsilon(\beta)}$ is the flux quanta through the qubit (coupling) loop. The circuit Hamiltonian is given by
\begin{equation} \label{eq:circ_ham}
    H_c = \frac{1}{2\phi_0^2}\pmb{p}^TC^{-1}\pmb{p}+U,
\end{equation}
where $C$ is the capacitance matrix of the system, $\pmb{p}$ is the momentum vector that can be described in terms of the charge number $p_i = \hbar n_i$ for each island of the superconducting circuit, and $U$ is the potential energy stored in the Josephson junctions. The potential energy is given by
\begingroup\makeatletter\def\f@size{9}\check@mathfonts
\def\maketag@@@#1{\hbox{\m@th\large\normalfont#1}}%
\begin{align}
\begin{split}
     U= -&\phi_0[I_{c1}\cos\gamma_1+I_{c2}\cos\gamma_2+I_{c3}\cos(\gamma_1+\gamma_2+\gamma_4+2\pi f_\epsilon)\\
    &+I_{c4}\cos\gamma_4+I_{c5}\cos\gamma_5+I_{c6}\cos(\gamma_4+\gamma_5-2\pi f_\beta)].
\end{split}
\end{align}\endgroup
The Hamiltonian in Eq.~(\ref{eq:circ_ham}) can be represented in the charge basis. A suitably truncated representation, with a large enough number of charge states, is used to numerically calculate relevant properties, including transition frequencies, matrix elements, and symmetry points.

When considering the interaction between the TLS and the transmission line, the relaxation rate reduces to
\begin{align}
    \Gamma_1 =\frac{\phi_0^2}{\hbar Z_{0}}\left|\gamma_{5, 10}\right|^{2}\Delta,
\end{align}
$Z_0=\sqrt{l_0/c_0}$ is the characteristic impedance of the transmission line and $\gamma_{5,10}=\bra{1}\gamma_5\ket{0}$ is the off-diagonal matrix element of the $\gamma_5$ phase operator between the ground state and the first excited state of the TLS (see Appendix~\ref{sec:SBFromCircuit} for details). By numerically solving Eq.~(\ref{eq:circ_ham}) and comparing the solution to Eq.~(\ref{eq:alpha}), the coupling strength for interaction in the system can be determined.

\section{Results}\label{sec:CalandResults}

\subsection{Device Characterization}

\begin{figure*}
    \centering
    \includegraphics{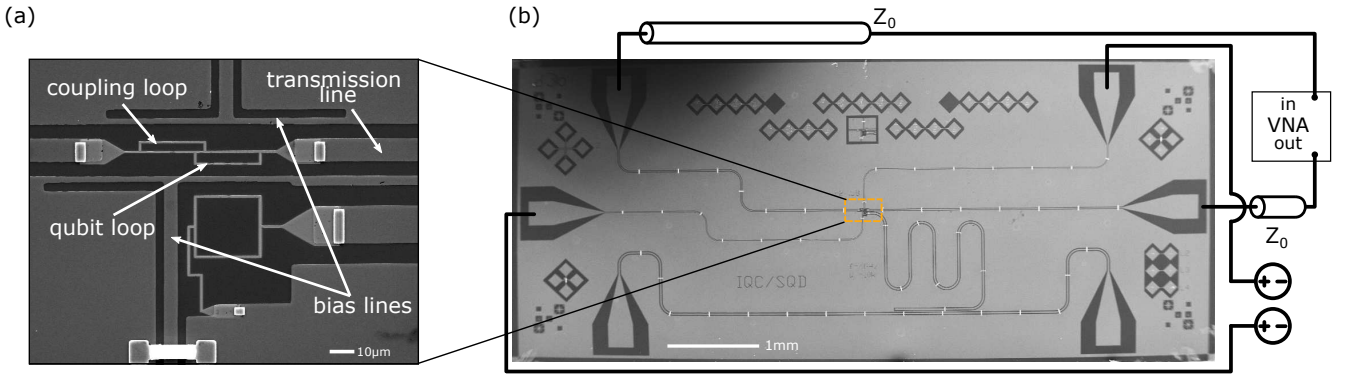}
    \caption[SEM images of Tunable Coupler Device]{(a) An SEM image showing the qubit, coupler, and bias lines. (b) An SEM image of the full device and a schematic representation of the experimental setup. The transmission line is connected to a VNA for measurement and the bias lines are connected to external voltage sources.}
    \label{fig:device}
\end{figure*}

A scanning electron microscope (SEM) image of the device, shown in Fig.~\ref{fig:device}(a), depicts the qubit loop, the coupling loop, and the TL of the circuit. The six Josephson junctions in the two loops are each set to have a specific critical current controlled through its junction area. With a target critical current density of 3~$\upmu \t{A}/\upmu \t{m}^2$, the junction areas are designed to be $A_1 = A_3 = 1.69A_0$, $A_2 = 0.58A_1$, $A_4 = A_5 = 3A_0$, and $A_6 = 0.52A_4$, where $A_0 = 0.0467~\upmu \t{m}^2$. The fluxes through the superconducting loops are controlled via two bias lines which are operated with DC currents to tune the coupling of the system. The bias lines are designed to be compatible with microwave pulses to fast-switch the coupling of the system as discussed in Appendix \ref{sec:sample}. The bottom line is designed to bias the qubit loop and interacts through a simulated mutual inductance of $M_\epsilon=0.25$~pH while the top bias line interacts with the coupler loop through a simulated mutual inductance $M_\beta=0.47$~pH. The device includes a DC-SQUID, which is inductively coupled to the qubit loop and can be used for readout of the TLS. The switching pulses and readout system were not required for this experiment. The full device and the schematic of setup are displayed in Fig.~\ref{fig:device}(b). The bias line currents are controlled by external voltage sources and the transmission through TL is measured with a vector network analyzer (VNA).

In our device, which has multiple bias lines and superconducting loops in close proximity, there is unavoidable flux crosstalk during operation. Measuring crosstalk is needed in order to enable independent control of $f_\beta$ and $f_\epsilon$. The system is calibrated by measuring the periodicity of the transmission through the TL at a single frequency versus the currents along the two bias lines. Voltage applied to each line generates a current that induces flux through both loops, which tunes the TLS transition frequency $\omega_{10}$ periodically. When the detuning, $\delta=\omega_p - \omega_{10}$, between the probe frequency and the qubit gap approaches zero, the photons are absorbed producing a visible dip in the transmission $|S_{21}|$. These dips form a closed contour when performing a sweep of $f_\beta$ and $f_\epsilon$. These contours occur near $(f_\beta,f_\epsilon)=(0.5+n,0.5+m)$ for integer values of $n$ and $m$.

\begin{figure}
    \centering
    \includegraphics[width=3.4in]{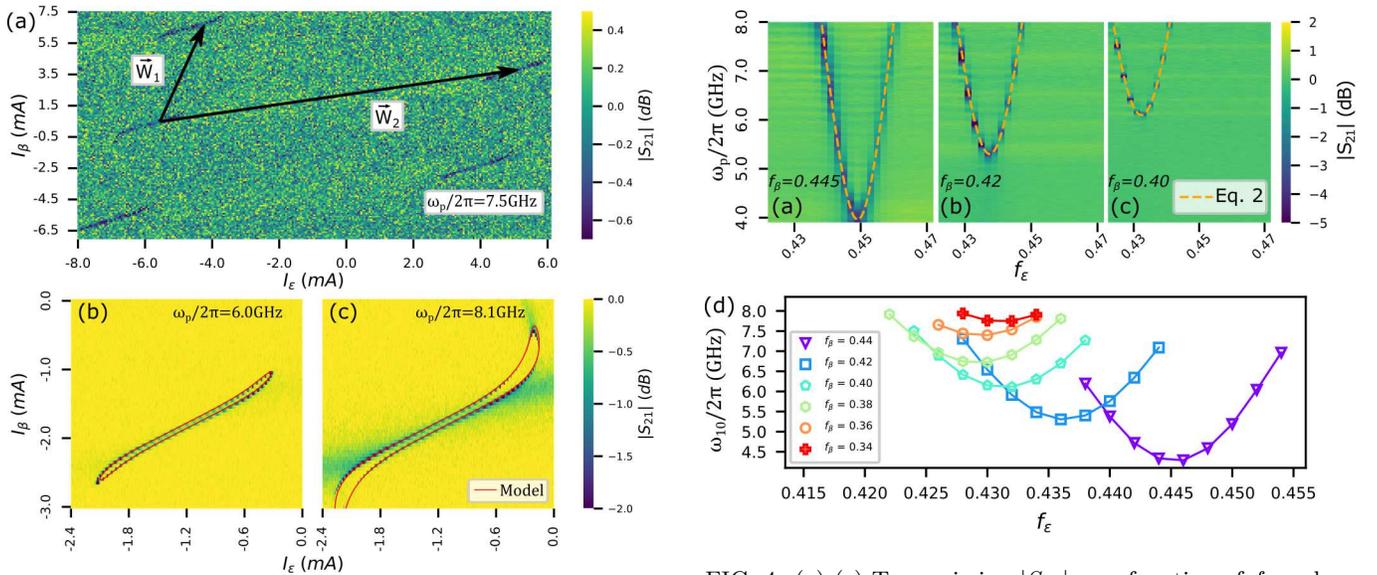}
    \caption{(a) The transmission $|S_{21}|$ as a function of $I_\epsilon$ and $I_\beta$ in the bias lines at $7.5~\text{GHz}$. Two translation vectors $\protect\overrightarrow{\boldsymbol{W_1}}$ and $\protect\overrightarrow{\boldsymbol{W_2}}$ identified based on periodicity. (b)-(c) Spectroscopy data when the probe frequency $\omega_p$ is set to $6.0,8.1~\text{GHz}$ respectively. The red line overlay shows the calculated corresponding contours based on the fitted circuit model.}
    \label{fig:Cal}
\end{figure}

The normalized fluxes in the loops are given by
\begin{gather}
 \begin{pmatrix} f_{\beta} \\ f_{\epsilon} \end{pmatrix}
 =
 \begin{pmatrix}
    W_{\beta\beta}& W_{\beta\epsilon} \\ W_{\epsilon\beta} & W_{\epsilon\epsilon}
\end{pmatrix}^{-1}
\begin{pmatrix}
   I_{\beta}-I_{0,\beta}  \\ I_{\epsilon}-I_{0,\epsilon} 
\end{pmatrix}
+ 
\begin{pmatrix}
   0.5 \\ 0.5
\end{pmatrix},
\end{gather}
where $W_{ij}$ are the elements of the crosstalk matrix, $I_{\epsilon(\beta)}$ is the applied current in the bottom (top) bias line, and $I_{0,\epsilon (\beta)}$ are the offset currents corresponding to the circuit biased at $(f_\beta,f_\epsilon)=(0.5,0.5)$. The crosstalk matrix is determined by scanning over a wide range of currents, and then using an image analysis routine to determine the periodicity of the data \cite{dai_calibration_2021}. In addition, image inversion symmetries are used to measure the current offset. Fig. \ref{fig:Cal}(a) shows the transmission through the TL for $\omega_p=7.5~\text{GHz}$ versus applied currents .We extract the translation vectors $\overrightarrow{\boldsymbol{W_1}}=(W_{\beta\beta},W_{\epsilon\beta})$ and $\overrightarrow{\boldsymbol{W_2}}=( W_{\beta\epsilon},W_{\epsilon\epsilon})$, which correspond to increasing the flux bias by one flux quantum in the qubit and coupling loop respectively. Fig. \ref{fig:Cal}(b)-(c) shows additional scans, with a different offset, that illustrate how the contours change with frequency. 

\begin{figure}
    \centering
    \includegraphics{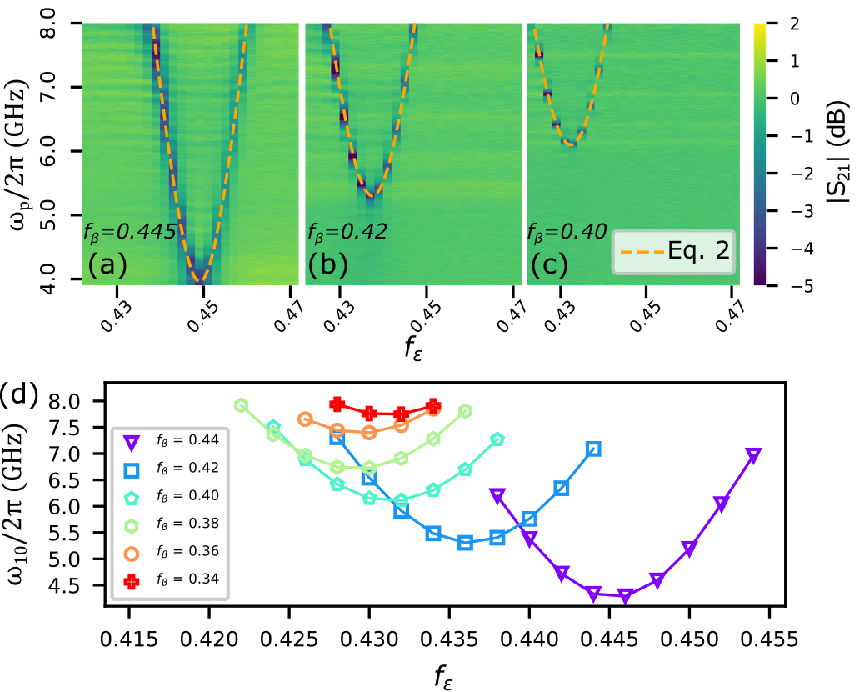}
    \caption[Transmission Spectroscopy for Various Flux Biasing in Coupling Loop]{(a)-(c) Transmission $|S_{21}|$ as a function of $f_\epsilon$ and $\omega_p$ at $f_\beta = 0.445, 0.42, 0.40$. The dashed line displays the fit to Eq. (\ref{eq:omega10}). (d) Data (symbols) and best fit model prediction (lines) for $\omega_{10}$ versus $f_\epsilon$  for values of $f_\beta$ ranging from 0.34-0.44.}
    \label{fig:Spectroscopy}
\end{figure}

In contrast to the case of standard flux qubits, $f_{\epsilon,\text{sym}}$ is not generally at $f_\epsilon=0.5$ due to an effective bias resulting from the neighbouring persistent current in the coupling loop. We can determine $f_{\epsilon,\text{sym}}$ for each $f_\beta$ by measuring the $f_\epsilon$ dependent qubit transition frequency from transmission spectroscopy and then fitting the Eq. (\ref{eq:omega10}) model as shown by Fig.~\ref{fig:Spectroscopy}(a)-(c). We note that the flux in the coupling loop affects the minimum transition energy $\Delta$ and, to a lesser extent, the persistent current $I_\t{TLS}$.

Using the numerical model developed in Sec.~\ref{sec:CircuitModel}, we fit the spectroscopically determined transition frequencies $\omega_{10}$ over a range of normalized fluxes  $f_\beta$ and $f_\epsilon$.  We make the assumption that the capacitance matrix can be accurately determined based on electromagnetic simulations, whereas Josephson junction critical currents $I_{c_i}$ are taken as fit parameters because they have more significant uncertainties arising from the fabrication process. As shown in Fig.~\ref{fig:Spectroscopy}(d), the best fit is in excellent agreement with the data, which is a compelling result given the relatively high complexity of the circuit, containing six Josephson junctions. Further validation of the model is provided by comparing the predicted transition frequency with the spectroscopy data taken during the calibration measurements (see Fig.~\ref{fig:Cal}(b)-(c)). 

Concerning the best fit parameters, we find the junction critical currents to be $I_{c_1}=0.236~\upmu \t{A}$, $I_{c_2}=0.131~\upmu \t{A}$, $I_{c_3}=0.236~\upmu \t{A}$, $I_{c_4}=0.411~\upmu \t{A}$, $I_{c_5}=0.584~\upmu \t{A}$, and $I_{c_6}=0.185~\upmu \t{A}$. Most of the junctions are in reasonable agreement to their design target with an error of $<12\%$, except for $I_{c_5}$, which was designed to be the same size as $I_{c_4}$ and has an error of approximately $40\%$. This is likely due to the inductive renormalization term neglected in the circuit model, which increases the effective critical current of junction 5 (see Appendix~\ref{sec:SBFromCircuit}).

\subsection{Tunable Coupling Strength}\label{sec:results}

To verify the tunability of the coupling, we characterize the transmission through the TL while setting the flux biases at symmetry points, where the transverse coupling is maximized. As shown in previous work~\cite{astafiev_resonance_2010,peropadre_scattering_2013,forn-diaz_ultrastrong_2017}, the transmission linewidth and the minimum transmission on resonance are related to the radiative loss from the qubit to the TL, as well as other loss channels coupled to the qubit. We extend previous work by including both the effects of finite temperature and large drive amplitude~\ref{sec:scatteringderivation}. With these extensions, the transmission is given by
\begin{align}
    t &=  \frac{1-r_0+\delta^{2} T_{2}{ }^{2}+2 T_{1} T_{2} \Gamma_{1} N_\t{in}+ir_0\delta T_2}{1+\delta^{2} T_{2}{ }^{2}+2 T_{1} T_{2} \Gamma_{1} N_\t{in}},\label{eq:t}
\end{align}
where $\delta = \omega_p-\Delta$ is the detuning of the probe frequency from the qubit gap, $T_{1(2)}$ is the qubit energy relaxation (dephasing) time due to all possible noise channels, $\Gamma_1$ is the relaxation rate to the TL in the zero temperature limit, and $N_\text{in}$ is the average number of incoming photons per second in the driving tone. The minimum transmission occurs when the probe frequency is on resonance and the incoming photon number $N_\t{in}$ is small, and is given by $1-r_0$ where
\begin{align}
    r_0 = \frac{1}{2} \frac{1-b}{1+b} T_{2} \Gamma_{1}.
\end{align}
Here $b=\exp(-\hbar\Delta/k_BT)$, where $T$ is the qubit effective temperature. It can be shown that $T_2 \rightarrow 2/\Gamma_1$ when the qubit temperature approaches zero and decoherence due to noises other than the TL becomes negligible and thus $r_0$ goes to 1 in this limit. Under the condition of strong coupling between the TLS and the TL and at a very low temperature, it is expected that we should reach this limit where the transmission through the TL goes to zero.

\begin{figure}
    \centering
    \includegraphics{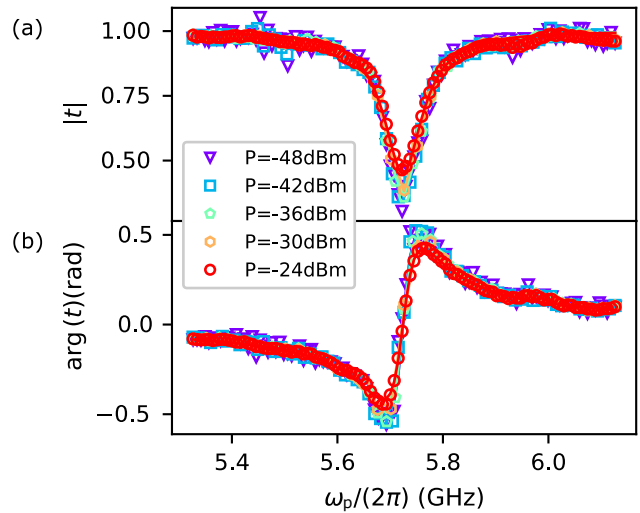}
    \caption[Transmission versus Power at Single Symmetry Point]{Transmission $t$ amplitude (a) and phase (b) versus probe frequency for powers ranging from -48~dBm to -24~dBm at a symmetry point $(f_\beta,f_\epsilon) = (0.41,0.433)$ where the qubit transition frequency $\Delta$ is $5.7~\text{GHz}$. The data is displayed as symbols and the best fit prediction based on the model using Eq. (\ref{eq:t}) is shown as lines.}
    \label{fig:PowerFit}
\end{figure}

The transmission data at the symmetry point for a range of $f_\beta$ values and powers $P$, is fit to Eq. (\ref{eq:t}) to extract $\Gamma_1$, as well as other relevant parameters. For each $f_\beta$, the transmission at different power is fitted simultaneously, with the rate of input photon given by $N_\t{in}= 10^{(P - A)/10-3}/\hbar\omega_p$, where $A$ is the attenuation. The parameters $T$ and $A$ for different $f_\beta$ are treated as independent parameters, and are found to be consistent with the estimates of the noise temperature from the bias line and the attenuation along the signal delivery chain (see Sec.~\ref{sec:Decoherence}). An example of the fit is shown in Fig. \ref{fig:PowerFit} for $(f_\beta,f_\epsilon) = (0.41,0.433)$.

\begin{figure*}
    \centering
    \includegraphics{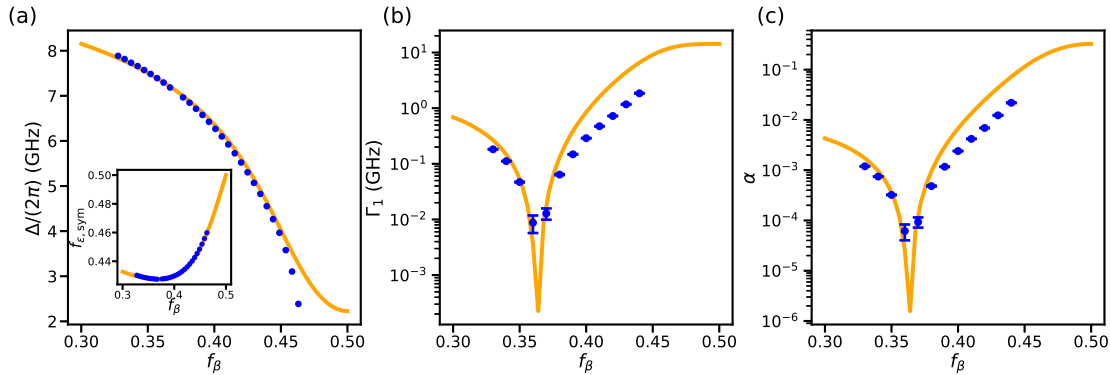}
    \caption[Coupling Results of Tunable Coupler System]{The extracted (dots) versus simulation (continuous lines) of key parameters for the coupled system. For each $f_\beta$ value, $f_\epsilon$ is set to the corresponding symmetry point.  (a) The qubit gap frequency. The inset shows values for $f_{\epsilon,\text{sym}}$. Values with $f_{\beta}\geq 0.45$ are determined by fitting spectroscopy data to Eq. (\ref{eq:omega10}) as the symmetry points cannot be directly measured with our current setup bandwidth. (b) The relaxation rate $\Gamma_{1}$ measured by fitting spectroscopy data using Eq. (\ref{eq:t}). (c) The coupling $\alpha$ between the TLS and the TL. The measured coupling ranges from $\alpha_\text{min}=6.2\times10^{-5}$ and $\alpha_\text{max}=2.19\times10^{-2}$.}
    \label{fig:Fits}
\end{figure*}

The coupling between the TL and the TLS is characterized and compared to the circuit model simulation. In Fig. \ref{fig:Fits}(a), plots for both the qubit splitting $\Delta$ and $f_{\epsilon,\text{sym}}$  versus $f_\beta$ are shown, which are predicted well by the circuit model. The zero temperature relaxation rate $\Gamma_{1}$, a measure of the coupling strength, is shown in Fig. \ref{fig:Fits}(b). Finally, the normalized coupling strength $\alpha$ is plotted in Fig. \ref{fig:Fits}(c) following Eq. (\ref{eq:alpha}). The minimum measured value of $\Gamma_{1}=8.7~\text{MHz}$ is at $f_{\beta}=0.36$ while the maximum measured value is $\Gamma_{1}=1.85~\text{GHz}$ at $f_{\beta}=0.44$. These produce a range of values for $\alpha$ where $\alpha_\text{min}=6.2\times10^{-5}$ and $\alpha_\text{max}=2.19\times10^{-2}$. It is not possible to measure the coupling for $f_{\beta}\geq0.45$ due to a limitation of the transmission bandwidth of the setup. Despite the discrepancy between the measured and simulated coupling, we still expect the maximum coupling to approach the USC regime, which is usually associated with  $\alpha \gtrsim 0.1$ \cite{forn-diaz_ultrastrong_2017,kockum_ultrastrong_2019}, by extrapolating the measured data to $f_{\beta}=0.5$. The circuit model also predicts that the minimum coupling is likely well below the smallest experimental value. At the decoupling point ($f_{\beta}\approx0.365$), the interaction strength is very small causing any dip in transmission to have a narrow width and become dominated by the residual dephasing, thus no smaller coupling is measured. This result demonstrates that we can effectively turn the coupling on or off by changing the flux bias through the coupling loop. 

\section{Discussion}\label{sec:discussion}

\subsection{Pure dephasing and residual relaxation}\label{sec:Decoherence}
\begin{figure}
    \centering
    \includegraphics[width=3.4in]{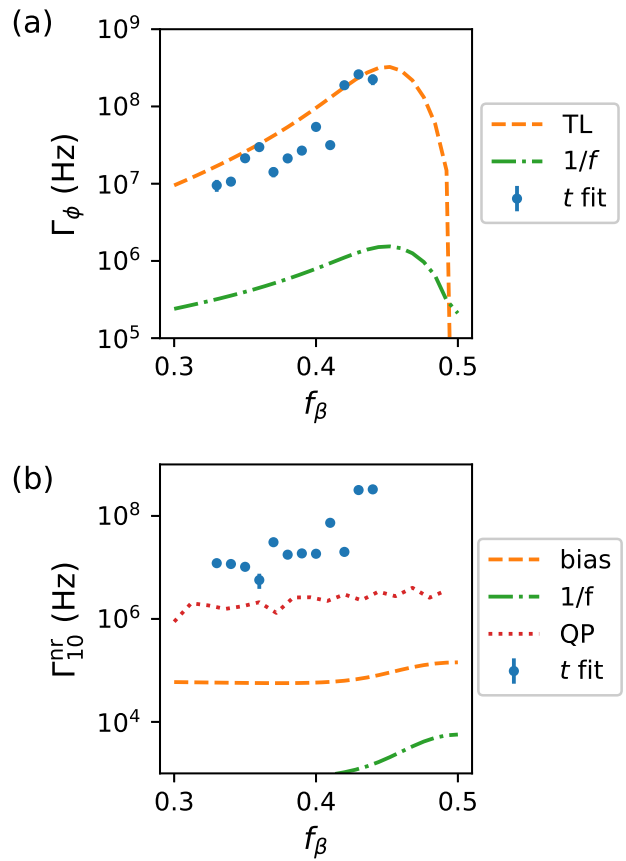}
    \caption{(Color Online) (a)Fitted pure dephasing rates (solid blue dots) and estimated pure dephasing rates due to the TL (orange dashed line) and $1/f$ flux noise (green dash-dotted line) versus normalized flux bias $f_\beta$. (b)Fitted (solid blue dots) residual relaxation rates and estimated relaxation rates due to the bias lines (orange dashed line), $1/f$ flux noise (green dash-dotted line) and quasiparticles (red dotted line), versus normalized flux bias $f_\beta$.}
    \label{fig:CoherenceEstimates}
\end{figure}

In this section we discuss the qubit relaxation and dephasing rates extracted from the transmission fit, as well as noise channels that could account for the rates. We first discuss the pure dephasing rates. As shown in Appendix~\ref{sec:SBFromCircuit}, the TL couples to the qubit longitudinally even when the qubit is at the symmetry point. Low frequency noise in the TL hence causes dephasing, with the dephasing rate given by
\begin{align}
    \Gamma_\phi^\text{TL}=\frac{\phi_0^2}{\hbar^2}|\gamma_{5,11}-\gamma_{5,00}|^2S_{I}(0)
\end{align}
where $S_{I}(0)$ is the current noise in the TL at DC. The current noise is given by the Johnson-Nyquist noise of the impedance $Z_0$ from the TL input, noting the output has a bandpass filter which is shorted to ground at DC. Therefore, the dephasing due to the TL is given by
\begin{align}
    \Gamma_\phi^{\t{TL}}=\frac{2k_BT\phi_0^2}{\hbar^2Z_0}|\gamma_{5,11}-\gamma_{5,00}|^2,
\end{align}
and it is plotted versus $f_\beta$ in Figure.~\ref{fig:CoherenceEstimates}(a). The estimated pure dephasing due to the TL agrees well with the dephasing rate found by the transmission fits. We note that pure dephasing due to the TL can be significantly reduced by adding a DC block, without changing relevant aspects of the TLS-waveguide interactions significantly.

Another important source of dephasing for flux tunable devices is the intrinsic $1/f$ flux noise. We assume the flux noise is of the form $A_{\epsilon,\beta}/(\omega/2\pi)$. The noise amplitudes are $\sqrt{A_\epsilon}=1.2~\mu\Phi_0/\sqrt{\text{Hz}},\sqrt{A_\beta}=1.1~\mu\Phi_0/\sqrt{\text{Hz}}$, estimated based on scaling the geometry of the loops compared to a previous device fabricated using similar process~\cite{orgiazziFluxQubitsPlanar2016}. The estimated dephasing due to $1/f$ flux noise is plotted in Figure.~\ref{fig:CoherenceEstimates}(a), and found to be much smaller than the dephasing extracted from the transmission fits.

We next discuss qubit relaxation, which we distinguish between radiative relaxation to the TL and non-radiative relaxation to all other noise sources. The qubit relaxation and excitation rates are separated into
\begin{align}
    \Gamma_{10(01)} = \Gamma_{10(01)}^\text{r}+\Gamma_{10(01)}^\t{nr}.
\end{align}
It needs to be noted that while the transmission line is likely thermalized at the dilution refrigerator base temperature, the bias lines have limited attenuations along the signal delivery chain and are likely to have much higher noise temperature. Hence the qubit effective temperature, extracted from the transmission fits, is a result of balancing the radiative and non-radiative thermalization rates and temperatures of the respective noise environment. We assume that the TL is in thermal equilibrium with the dilution refrigerator, at temperature $T^\text{TL}=50~\text{mK}$. This gives
\begin{align}
    \Gamma_{10}^\text{r}=\Gamma_1\frac{\exp{(\hbar\omega_{01}/k_BT^\text{TL})}}{\exp{(\hbar\omega_{01}/k_BT^{TL}})-1},\\
    \Gamma_{01}^\text{r}=\Gamma_1\frac{1}{\exp{(\hbar\omega_{01}/k_BT^{\text{TL}}})-1}.
\end{align}
Assuming the noise channels other than the TL is also in thermal equilibrium at some temperature $T^\t{nr}$, we have 
\begin{align}
    \frac{\Gamma_{01}^\t{nr}}{\Gamma_{10}^{nr}}=\exp \left(\frac{-\hbar\omega_{01}}{k_BT^\t{nr}}\right)
\end{align}
The extracted residual relaxation $\Gamma_{10}^\t{nr}$ is shown in Fig.~\ref{fig:CoherenceEstimates}(b). We note that the the residual noise temperatures found by the transmission fits is on the order of hundreds of $\text{mK}$, consistent with the estimated noise temperature based on the attenuation applied along the bias lines.

We next consider the possible causes of the residual relaxation. One important relaxation channel is the flux bias lines used to enable independent biasing of the two loops in the device. The relaxation rate is proportional to the square of the mutual coupling between bias lines and the normalized flux bias matrix elements $\langle0|\partial H/\partial f_\epsilon |1\rangle, \langle0|\partial H/\partial f_\beta |1\rangle$. Using the mutual inductance values found from the calibration measurement, the calculated relaxation rate due to bias line thermal noise is shown in Fig.~\ref{fig:CoherenceEstimates}(b). It can be seen that this noise channel introduces relaxation rate of about $100~\text{kHz}$, much smaller than the relaxation found from the transmission fits. 

Another possible source of relaxation is intrinsic flux noise, which in general follows the $1/f$ spectrum up to a few $\text{GHz}$. Using the same noise amplitudes used for the dephasing estimates, we found the relaxation due to intrinsic flux noise below $10~\text{kHz}$, which is negligible. 

Finally we consider relaxation due to quasiparticle tunneling across the junctions. Following Ref.~\cite{catelaniQuasiparticleRelaxationSuperconducting2011}, the relaxation rate is given by
\begin{align}
    \Gamma_{10}^{\t{qp}}=\sum_{i=1}^6\left|\bra{0}\sin\frac{\gamma_i}{2}\ket{1}\right|^2\frac{8x_{\text{qp}} E_{J_i}}{\hbar\pi}\left(\frac{2\Delta_\text{Al}}{\hbar\omega}\right)^{\frac{1}{2}},
\end{align}
where $x_{qp}$ is the quasiparticle density normalized by the density of superconducting electrons, $E_{J_i}=I_{c_i}\phi_0$ is the Josephson energy of junction $i$ and $\Delta_{\text{Al}}$ is the superconducting gap of Al. Assuming a modest value of the normalized quasiparticle density $x_{\text{qp}}=5\times10^{-7}$\cite{yan_2016_fluxqubitrevisited,paikObservationHighCoherence2011} on all qubit islands, and using the tunneling matrix elements $\bra{0}\sin(\gamma_i/2)\ket{1}$ on all six junctions computed from the circuit model, we found that the quasiparticle tunneling induced relaxation is about an order of magnitude smaller than the residual relaxation found from fitting the qubit transmission data (see Figure.~\ref{fig:CoherenceEstimates}(b)). However, it is possible that our device has a much larger quasiparticle density, due to the relatively high noise temperature on the bias lines or other sources. We expect future iteration of the experiments would have improved signal delivery setup, which could reduce the noise temperature and hence improve the qubit relaxation times. 

\subsection{Investigating Discrepancy at Large Coupling}\label{sec:discrepancy}

In this section, we investigate potential explanations for the observed discrepancy between the measured and predicted coupling at large coupling strength. We first consider the effect of qubit renormalization due to the transmission line. The renormalization term, typically neglected in literature~\cite{peropadre_nonequilibrium_2013}, adds an additional inductive potential term to the potential energy of the qubit-coupler circuit that could cause a deviation in the coupling of the system. However, after numerical simulation, we find the effect of the renormalization term can be mostly captured by an increase in the fitted junction 5 parameter with little effect on the coupling (see Appendix~\ref{sec:SBFromCircuit}).

Next we consider the effects of finite reflections in the transmission line as a possible cause for the discrepancy. As shown in Appendix~\ref{sec:SBFromCircuit}, the coupling between the TLS and transmission line can be described by the spin-boson model. This model assumes an infinite, ideal transmission line, but finite reflections in our setup at microwave filters around the device could cause a deviation from the model. To investigate this, we model the filter as a reciprocal, lossless component with reflections characterized by a voltage standing wave ratio (VSWR) in a microwave network. Following the canonical quantization procedure, we find that, due to the finite reflections, modes in the transmission line are modified to be superpositions of forward and backward traveling waves. Interference at the position of the qubit then leads to variation in the current fluctuations of the transmission line  seen by the qubit, thus modulating the coupling strength at different frequencies. For further details of this derivation, see Appendix~\ref{sec:reflections}. 

\begin{figure}
    \centering
    \includegraphics[]{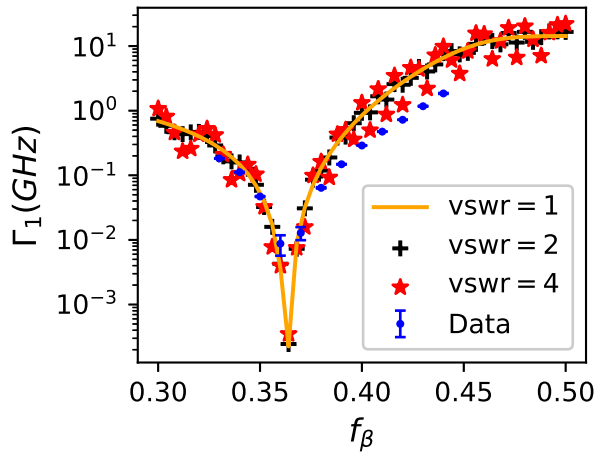}
    \caption{The measured (blue dots with error bar) and predicted radiative relaxation rate $\Gamma_1$ using Eq.~\ref{eq:FilterRelaxation} assuming VSWR=$1$(orange line), $2$(black crosses) and $4$ (red stars).}
    \label{fig:GammaFilter}
\end{figure}

Using this new model and assigning an arbitrary VSWR to the filter component we can observe the effects of finite reflections on the coupling of the system. We calculate the radiative relaxation rate $\Gamma_1$ as a function of $f_\beta$, for VSWR~$=1,2,4$. As shown in Fig.~\ref{fig:GammaFilter}, increasing reflection leads to oscillations in $\Gamma_1$ with increasing amplitudes around the predicted values in the no reflection limit (VSWR~$=1$). Taking the lower values in the oscillation, $\t{VSWR}\gtrsim 4$ is required to explain the discrepancy between measured values compared to the theoretical values in the no reflection limit. Room temperature characterization of the measurement setup indicate that the VSWR due to components on the output of the transmission line is roughly $1.5$, smaller than the value required to explain the inconsistency. Further analysis of the impact and potential causes for the discrepancy in the coupling strength will be pursued in future work.

\section{Summary and Outlook}\label{sec:summary}

In summary, we propose and implement a superconducting device that enables the tunable coupling of a TLS to a waveguide over several orders of magnitude of coupling strength. After calibrating for flux crosstalk, a circuit model was fit to the spectroscopic which showed very good agreement. The coupler demonstrated light-matter interaction strength from a decoupled state to near the USC regime with our model predicting a greater achievable range. We investigate potential causes for an observed discrepancy between the measured and predicted coupling at large coupling strength including reflections in the transmission line and the addition of a renormalization term in the spin-boson model, but do not find a complete explanation in this work.

The tunable coupler enables several avenues for future research using the wide range of coupling strength and the ability to effectively decouple the TLS from the transmission line. Fast switching can be used to directly measure the light-matter time dynamics in the USC regime by implementing experiments proposed in Ref. \cite{shi_fast_2019}. With further optimization of the design parameters we expect to be able to increase the maximum coupling and reach coupling strengths well into the USC regime. Further, the device has application in photon packet production where the TLS can be excited and then rapidly coupled to the TL to emit shaped single photons \cite{houck_generating_2007,forn-diaz_-demand_2017}. Finally, the device can be used for relativistic quantum information experiments where two of the couplers allow spatially separated qubits to independently interact with a field over very short time scales to demonstrate entanglement harvesting from a vacuum \cite{salton_acceleration-assisted_2015,pozas-kerstjens_harvesting_2015}.

\begin{acknowledgments}
\section*{Acknowlegments}

 We are grateful to Ali Yurtalan for advice during the early stages of the project. We acknowledge support from NSERC through Discovery and RTI grants, Canada Foundation for Innovation, Ontario Ministry of Research and Innovation, and Industry Canada. We would like to acknowledge support from CMC Microsystems and Canada’s National Design Network (CNDN). The University of Waterloo's QNFCF facility was used for this work. This infrastructure would not be possible without the significant contributions of CFREF-TQT, CFI, ISED, the Ontario Ministry of Research \& Innovation and Mike \& Ophelia Lazaridis.
 
\end{acknowledgments}

\appendix

\section{Sample Preparation and Measurement Setup}\label{sec:sample}
The device is fabricated using a process that includes 3 aluminum layers consisting of waveguides, Josephson Junctions, and bridges/bandages respectively. The waveguide layer is patterned using negative photo resist on a silicon wafer, followed by evaporating $100~\t{nm}$ of aluminum.  The junction layer, which includes the qubit and the coupling loops, is made using standard double angle shadow evaporation with two aluminium depositions of thickness $40~\t{nm}$ and $70~\t{nm}$ respectively. The final layer serves a dual purpose of creating air bridges to act as ground plane interconnects as well as a layer for generating bandages to ensure galvanic contact between the circuit layer and the junction layer. For this step, argon milling is first used to galvanically connect the deposited aluminium to the previous layers. Then, $450~\t{nm}$ of aluminium is deposited on PMMA scaffolds created with gray scale e-beam lithography~\cite{janzen_aluminum_2022}.

The device is mounted in a copper sample box placed at the mixing chamber plate of a dilution refrigerator unit with a base temperature of $30~\t{mK}$. The sample is placed inside a magnetic shield formed of three layers of paramagnetic material. Microwave components near the sample are chosen to be non-magnetic to minimize potential noise and offset magnetic flux at the device. The received from the device along the transmission line (TL) are amplified using a $4-8~\t{GHz}$ amplifier, which limits the coupling measurements performed on this device to $\Delta$ values within this bandwidth, or $f_{\beta}<0.45$. The signals sent to and received from the device are each filtered through a $4-8~\t{GHz}$ band pass filter placed near the sample to reduce quasi particle noise from outside this bandwidth. The TL is connected to an Agilent E5071C vector network analyser for transmission measurements.

The on chip bias lines are designed with the goal of fast-switching the interactions between the qubit and the transmission line. This is achieved by designing the on chip waveguide feeding the flux bias lines with an impedance of 50 ohm, matching external transmission lines. In addition, Air bridges prevent mode conversion between waveguide and stripline modes. Finally, the mutual inductances from the bias lines to the qubit loops, together with the filtering and attenuations on the microwave lines, are designed to allow a large range of flux tunability while minimizing the negative impact on qubit coherence (see Sec.~\ref{sec:Decoherence}). With appropriate protocols, the fast lines can be operated with pulses as short as ${\sim}100~\t{ps}$ to fast-switch the coupling in and out of the USC regime. The coupler design also includes a DC-SQUID coupled to a resonator for readout of the TLS. We note that while the pulsed mode operation and readout are not needed in the current experiment, which focuses on the measurement of the coupling strength, they will be relevant for future experiments that involve time-domain measurements of the system.

In addition to microwave signals, the bias lines are also connected to DC sources, which are combined using a bias-tee placed at the mixing chamber of the dilution refrigerator. The DC lines are connected to Yokogawa 7651 external power supplies which supply the bias current for this experiment. The currents applied through the bias lines are larger than initially anticipated, due to misestimates of the mutual inductances and the malfunctioning of a global flux control. This led to significant heating of the DR during operation, with temperature reaching ${\sim}50~\t{mK}$ during the transmission measurements. This heating effect can be prevented in future work by replacing the bias wires with superconducting wires


\section{Spin-boson Model for Superconducting Circuits}\label{sec:SBFromCircuit}
In this section, we discuss details of the derivation of the coupling between a two-level system and a transmission line, and its correspondence with the spin-boson model \cite{leggett_dynamics_1987}. A description of this form has been used in previous literature \cite{peropadre_nonequilibrium_2013,forn-diaz_ultrastrong_2017,shi_fast_2019}.

The spin-boson model Hamiltonian is given by
\begin{align}
   H_\text{SB}&=H_{s}+H_{b}+H_\text{int},
\end{align}
where $H_{s}$ is the Hamiltonian for an atom implemented as a TLS, $H_{b}$ is the Hamiltonian for a bath of bosonic modes, and $H_\text{int}$ is the Hamiltonian describing their interaction. We have
\begin{align}
    H_{s} &= -\frac{\hbar\omega_{10}}{2}\sigma_z,\\
    H_b &=\sum_k \hbar\omega_k a_k^\dagger a_k~\text{and}\\
    H_\text{int} &=\sum_k (g_k^x\sigma_x+g_k^z\sigma_z) (a_k^\dagger+a_k),
\end{align}
where $\omega_{10}$ is the transition frequency of the TLS, $\sigma_z$ and $\sigma_x$ are the system's Pauli operators in the energy eigenbasis, $g_k^x, g_k^z$ are the transverse and longitudinal coupling strength of the $k^{\text{th}}$ mode of the bosonic bath, and $\omega_k$, $a_k{}^\dagger$, $a_k$ are frequency, creation, and annihilation operators of the $k^{\text{th}}$ mode of the bosonic bath.

In the Born-Markov limit, the interaction leads to qubit relaxation with the rate
\begin{align}
    \Gamma_1 = \frac{2\pi}{\hbar^2}\sum_kg_k^2\delta(\omega -\omega_k)=J(\omega)\label{eqn:SBGamma},
\end{align}
where $\Gamma_1$ is the relaxation rate when the bath temperature is zero, which we also associate with the bath spectral density $J(\omega)$. For an ohmic bath, a case that applies to our open transmission line, $J(\omega)$ can be written in terms of the dimensionless coupling constant $\alpha_{\text{SB}}$ 
\begin{align}
    J(\omega)=\pi\alpha_\t{SB}\omega.
\end{align}

The spin-boson model is realized with a superconducting flux qubit tunably coupled to an open transmission line. The Hamiltonian is given by
\begin{align}
    H=\int_{-L/2}^{L/2}\left[\frac{q(x)^{2}}{2 c_{0}}+\frac{\partial_{x} \Phi(x)^{2}}{2 l_{0}}\right]\mathrm{d} x + H_{\mathrm{c}}+H_c^\prime+H_\t{int}.
\end{align}
Here $q(x)$ and $\Phi(x)$ are the charge and phase at position $x$ along the line, $c_0$ and $l_0$ are the characteristic capacitance and inductance of the line, and $L$ is the length of the TL. The component $H_c$ is the circuit Hamiltonian consisting of the capacitive and Josephson energies in the coupling and qubit loops, as described in Sec.~\ref{sec:CircuitModel}. The component $H_c^\prime$ is a renormalization term acting on the coupler-qubit circuit due to the transmission line,
\begin{align}
    H_c^\prime=\frac{\phi_0^2\gamma_5^2}{2l_0\delta x},
\end{align}
where $\gamma_5$ is the Josephson phase of junction 5, the coupling junction, and $\delta x$ is a relevant length scale at which the transmission line can be modeled as a chain of discrete LC circuits. Following~\cite{peropadre_nonequilibrium_2013}, $\delta x$ can be taken to be on the order of $v/(\omega_{10}/2\pi)$, with $v=\sqrt{1/(l_0c_0)}$ the speed of light in the TL. The renormalization term $H_c^\prime$ can in principle be absorbed into the circuit Hamiltonian $H_c$. Through numerical simulations, it is found that the inclusion of $H_c^\prime$ affects the model very similar to a simple increase junction 5's critical current, resulting in nearly identical qubit transition frequencies and coupling strengths. Hence the term is neglected in the coupler-qubit circuit model fitted in this work.

The interaction Hamiltonian is given by
\begin{align}
    H_\t{int}&=\frac{1}{l_{0}} \gamma_5\phi_0\partial_{x} \Phi(x=0),
\end{align}
which describes the coupling between the harmonic modes and the tunable coupler circuit via the Josephson phase of the coupling junction $\gamma_5$ assuming the coupler circuit is at $x=0$. 

The field in the TL can be quantized and the interaction is given by ~\cite{forn-diaz_ultrastrong_2017}
\begin{align}\label{eqn:CircuitInt}
    H_{\mathrm{int}}=\frac{1}{l_{0}} \phi_{0} \gamma_{5} \sum_{k} \sqrt{\frac{\hbar}{2 c_{0} \omega_{k} L}} &k\left(i a_{k} e^{i\left(k x-\omega_{k} t\right)}\right.\\\nonumber
    &\left.-i a_{k}^{\dagger} e^{-i\left(k x-\omega_{k} t\right)}\right),
\end{align}
where angular wavenumber $k$ is related to the frequency via $k=\omega_k/v$. Comparing Eq.~(\ref{eqn:CircuitInt}) with the spin-boson interaction Hamiltonian the transverse coupling constant is obtained as
\begin{align}
    g_{k}^x=\frac{1}{l_{0}} \phi_{0}\left|\gamma_{5, 10}\right| \frac{1}{\sqrt{L}} \sqrt{\frac{\hbar \omega_{k}}{2 c_{0} v^{2}}},
\end{align}
where $\gamma_{5, 10}=\bra{1}\gamma_5\ket{0}$ is the off-diagonal matrix element between the first excited state and the ground state of the $\gamma_5$ phase operator. Finally, to obtain the spectral density or equivalently the relaxation rate at the symmetry point, one takes the limit $L\xrightarrow[]{}\infty$, replaces the summation in Eq.~(\ref{eqn:SBGamma}) with an integral (see Supplementary information of Ref.~\cite{forn-diaz_ultrastrong_2017}), and uses the minimum qubit transition frequency $\Delta$, which gives
\begin{align}
    \Gamma_1 =\frac{\phi_0^2}{\hbar Z_{0}}\left|\gamma_{5, 10}\right|^{2}\Delta,
\end{align}
where $Z_0=\sqrt{l_0/c_0}$ is the characteristic impedance of the line. 

\begin{figure}
    \centering
    \includegraphics{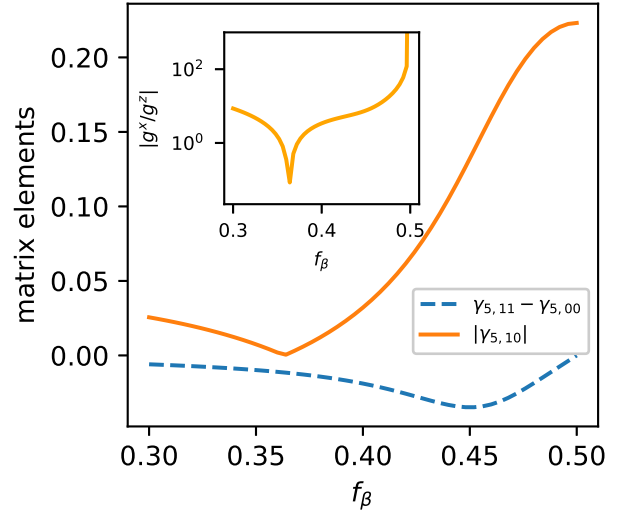}
    \caption{The matrix elements of $\gamma_5$ and the ratio of transverse and longitudinal coupling (inset) as a function of $\beta$-loop bias.  }
    \label{fig:x_vs_z}
\end{figure}

Besides the transverse coupling, the coupling loop also mediates a longitudinal coupling between the qubit and the TL, even when the qubit is at the symmetry point. The ratio of the coupling strengths satisfies
\begin{align}
    \left|\frac{g_x^k}{g_z^k}\right|&=\frac{2|\gamma_{5,10}|}{\left|\gamma_{5,11}-\gamma_{5,00}\right|},
\end{align}
and is plotted in Fig.~\ref{fig:x_vs_z}. It can be seen that the transverse coupling strength $g_x^k$ dominates except near $f_\beta\approx 0.365$, where $g^x_k\xrightarrow[]{}0$. We therefore expect the longitudinal coupling to have small influence on the tunable coupler device when operated in the strong to ultra-strong coupling limit. However, we note that this longitudinal coupling leads to pure dephasing, which we discuss in more detail in Sec.~\ref{sec:Decoherence}. 

\section{Influence of finite temperature and residual decoherence on the scattering parameter}\label{sec:scatteringderivation}
Our expression for the transmission can be obtained by extending the derivation in Ref.~\cite{peropadre_scattering_2013} to include explicitly the temperature of the qubit, as well as thermalization and dephasing due to noise channels other than the TL. 

Following Eq.~(47, 50) of Ref.\cite{peropadre_scattering_2013}, we write the input and output field from the left of the of the qubit as
\begin{align}
    V_\t{L}^\t{in} &=\Omega_p \sin{(\omega_p t)},\\
    V_\t{L}^\t{out} &=\frac{1}{2}\sqrt{Z_0\hbar\omega_{10}\Gamma_1}\langle{}{\sigma_x}\rangle,
\end{align}
where $\omega_p$ is the drive frequency and $\langle{}{\sigma_x}\rangle$ is the expectation value of the Pauli operator.
The input field acts as a drive on the qubit, inducing Rabi oscillations with frequency $\omega_r$. The Rabi frequency can be related to the input voltage amplitude $\Omega_p$ via the coupling strength between the TL and the TLS in terms of $\Gamma_1$, given by
\begin{align}
    \omega_r &=\frac{\Omega_p\phi_0}{\hbar Z_0}|\gamma_{5,10}|\\
    &= \Omega_p\sqrt{\frac{\Gamma_1}{Z_0\hbar\omega_{10}}}.
\end{align}
Based on Ref.~\cite{peropadre_scattering_2013}, the finite temperature and residual decoherence only affects the output field via a change of the equilibrium state of the qubit only. The equilibrium state of the qubit can be found by solving a generic master equation including relaxation, dephasing and a coherent drive on the qubit. The master equation is given by
\begin{align}
    \dot{\rho}&=-\frac{i}{\hbar}\left[ H_\text{s}+ H_\text{drive}, \rho\right]\\
    &+\Gamma_{01} D\left(\sigma_{-}\right) \rho+\Gamma_{10} D\left(\sigma_{+}\right) \rho+\frac{1}{2} \Gamma_{\phi} D\left(\sigma_{z}\right) \rho,
\end{align}
where $\Gamma_{01}$, $\Gamma_{10}$, and  $\Gamma_{\phi}$ are the noise induced excitation, relaxation, and pure dephasing rates respectively due to all sources of noise including coupling to the open TL, and $D(\sigma_{+})$, $D(\sigma_{-})$, and $D(\sigma_{z})$ are the corresponding Lindblad-form super-operator. We also introduce $\Gamma_{01}/\Gamma_{10}=b=\exp(-\hbar\omega_{10}/k_BT)$, where $T$ is the qubit effective temperature, defined in terms of the detailed balance of the total relaxation and excitation rates. The system Hamiltonian is given by
\begin{align}
    H_\text{s}&=-\frac{\hbar\omega_{10}}{2}\sigma_z,
\end{align}
and the drive Hamiltonian is 
\begin{align}
    H_\text{drive}&=\hbar\omega_r\sigma_y \sin{\omega_p t}.
\end{align}
Employing the rotating wave approximation we can find the steady-state solution of the master equation. In particular, the $\sigma_x$ expectation value is given by
\begin{align}
    \langle \sigma_x\rangle&=\frac{b-1}{1+b}\frac{T_2\omega_r}{1+\delta^2T_2^2+T_1T_2\omega_r^2}(\sin{\omega_p t}-\delta T_2\cos{\omega_p t}),
\end{align}
where $\delta$ is the detuning $\omega_p-\omega_{10}$ and we have defined the total relaxation and dephasing time, given respectively by
\begin{align}
    T_1&=\frac{1}{\Gamma_{10}+\Gamma_{01}}~\t{and}\\
    T_2&=\frac{1}{\Gamma_\phi+1/(2T_1)}.
\end{align} 
Finally the reflection coefficient is given by
\begin{align}
    r =  -r_0\frac{1-i \delta T_{2}}{1+\delta^{2} T_{2}{ }^{2}+2 T_{1} T_{2} \Gamma_{1} N_\text{in}},\label{eqn:Reflection}
\end{align}
where we have defined the average number of incoming photons
\begin{align}
    N_\t{in} &=\frac{\Omega_p^2}{2Z_0\hbar\omega_{10}},
\end{align}
and 
\begin{align}
    r_0 &= \frac{1}{2} \frac{1-b}{b+1} T_{2} \Gamma_{1}.
\end{align}
Substituting Eq.~(\ref{eqn:Reflection}) into $t=1+r$ allows us to obtain the transmission coefficient $t$, which is used to fit the measured $S_{21}$ after removing the background transmission. We note that in the zero temperature limit, the expression for the reflection Eq.~(\ref{eqn:Reflection}) reduces to the form derived in Ref.~\cite{peropadre_scattering_2013} (see Eq.~(53) of the reference).

\section{Transmission line modes with finite reflection}\label{sec:reflections}

As discussed in Appendix~\ref{sec:SBFromCircuit}, the spin-boson model can be obtained from the system of flux qubit coupled to the TL. The coupling strength is obtained based on the assumption of an infinite TL. This assumption is valid when there is no reflection in the TL, or the signal becomes significantly attenuated before reflection occurs. In our experiment setup, the filters on the output line can cause significant reflection. In this section, we discuss how finite reflection affects the TL modes and estimate the reflection required to explain the discrepancy of $\Gamma_1$ presented in the main text. 

\begin{figure}
    \centering
    \includegraphics[]{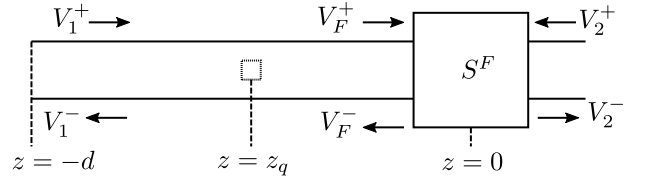}
    \caption{Schematic of the qubit embedded in a TL, where one side of the TL has non-negligible reflection, characterized by the scattering matrix $S^F$.}
    \label{fig:TLFilter}
\end{figure}

We model the filters and the entire microwave signal chain behind it as a generic component labelled as $\text{F}$, with scattering parameter $S^\t{F}$, placed at $z=0$ on the TL. The qubit is placed at position $z=z_q$. In Fig.~\ref{fig:TLFilter} the schematic of this model is shown, with TL length $d$. To simplify the analysis we will also assume that the component $F$ is reciprocal and lossless. 

The scattering parameter of the full system is related to $S^\t{F}$ by a shift of reference plane, given by
\begin{align}
    S(\omega) &=\begin{pmatrix}
    e^{-2i\omega d/v}S^\t{F}_{11}&e^{-i\omega d/v}S^\t{F}_{12}\\
    e^{-i\omega d/v }S^{\t{F}}_{21}&S^{\t{F}}_{22}
    \end{pmatrix},
\end{align}
where $\omega$ the angular frequency and $v$ the speed of light. 

In line with the periodic boundary condition, the allowed modes need to satisfy $V_1^+=V_2^-, V_1^-=V_2^+$. This leads to
\begin{align}
    \begin{pmatrix}
    V_1^-\\
    V_2^-
    \end{pmatrix}
    &=S(\omega)\begin{pmatrix}
    V_1^+\\
    V_2^+
    \end{pmatrix},\\
    \begin{pmatrix}
    V_1^-\\
    V_1^+
    \end{pmatrix}
    &=S(\omega)\begin{pmatrix}
    V_1^+\\
    V_1^-
    \end{pmatrix},
\end{align}
and the allowed frequencies need to satisfy
\begin{align}
    \t{det}\left[S(\omega)-\begin{pmatrix}
    0&1\\
    1&0
    \end{pmatrix}\right]=0.\label{eqn:FilterPBC}
\end{align}
Solving the above equation gives the allowed frequencies 
\begin{align}
    \omega_{s,n}&=\frac{\theta v}{d}+(-1)^s\frac{v}{d}2\pi n
\end{align}
where $n$ is an integer, $s\in{0,1}$ indicates the parity of the mode, and $\theta$ is a factor which goes to zero when there is no reflection at $F$. The allowed modes are in general superpositions of left and right travelling waves with their amplitudes satisfying
\begin{align}
    S_{11}(\omega_s)V_{s,1}^++\left(S_{12}(\omega_s)-1\right)V_{s,1}^-&=0,\\
    \frac{V_{s,1}^-}{V_{s,1}^+}=\frac{S_{11}(\omega_s)}{1-S_{12}(\omega_s)}\label{eqn:WaveRatio}.
\end{align}
To simplify notation, the subscript $n$ is dropped unless specified otherwise, since the values of the scattering matrix elements are independent of $n$. The density of states in the finite reflection case is unchanged from the zero reflection case, given by $2\pi v/d$ for each parity.

Since we will be taking the $d\xrightarrow[]{}\infty$ limit, and we are ultimately interested in the current fluctuation at the position of the qubit, it is useful to relate the ratio of left and right travelling voltage amplitudes at $z=-d$ and at just at the left of the component $F$. 
\begin{align}
    \frac{V_{s,F}^-}{V_{s,F}^+}&=S^F_{11}+\frac{S^F_{12}S^F_{21}}{\frac{1-S_{12}(\omega_s)}{S_{11}(\omega_s)}-S^F_{22}}=r_{s},
\end{align}
where $r_s$ indicates that this reflection constant only depends on the parity and not the mode number. Then the voltages and current on the TL for a particular mode can be written as
\begin{align}
    V_s(z,t)&=V_{s,F}^+\l\{\exp{\left[i\omega_s(t-z/v)\right]}\r.\nonumber\\
    &\l.+r_s\exp{\l[i\omega_s(t+z/v)\r]}\r\}+\text{\t{c.c.}},\label{eqn:Voltage}\\
    I_s(z,t)&=i\frac{V_{s,F}^+}{Z_0}\l\{\exp{\left[i\omega_s(t-z/v)\right]}\r.\nonumber\\
    &\l.-r_s\exp{\l[i\omega_s(t+z/v)\r]}\r\} +\text{\t{c.c.}},\label{eqn:Current}
\end{align}
where $\text{c.c.}$ stands for complex conjugate, added to ensure the voltage and current on the line is real. To quantize the field on the TL, we first consider the total energy stored in the line, given by
\begin{align}
    H_s&=\int_{-d}^0\l[\frac{c_0}{2}V_s(z,t)^2+\frac{1}{2l_0}I_s(z,t)^2\r]dz,\label{eqn:TLEnergy}\\
    &=2c_0d|V_{s,f}^+|^2(1+|r_s|^2),
\end{align}
where in the first equality assumes that component $F$ does not store energy. This is valid anticipating that we will take the $d\xrightarrow[]{}\infty$ limit, where the energy at any particular point on the line approaches zero. In the second equality, we discard terms oscillatory in $z$ as they too approaches zero in the $d\xrightarrow[]{}\infty$ limit. The energy expression Eq.~(\ref{eqn:TLEnergy}) can be related to the Hamiltonian of a harmonic oscillator by introducing the dynamical variables
\begin{align}
    q_s&=N_s(\widetilde{V^+_{s,F}}+\widetilde{V^{+}_{s,F}}^*),\\
    p_s&=\dot{q}_s=-i\omega_s N_s(\widetilde{V^+_{s,F}}-\widetilde{V^{+}_{s,F}}^*),
\end{align}
where $\widetilde{V^{+(-)}_{s,F}}=V^{+(-)}_{s,F}\exp{(i\omega_s t)}$ and $N_s$ is some normalization constant. By setting 
\begin{align}
    N_s=\sqrt{\frac{c_0d(1+|r_s|^2)}{\omega_s^2}},
\end{align}
the energy expression in Eq.~(\ref{eqn:TLEnergy}) becomes
\begin{align}
    H_s&=\frac{1}{2}p_s^2+\frac{1}{2}\omega_s^2q_s^2.\label{eqn:TLHam}
\end{align}
We note that the equations of motion of $q_s, p_s$ satisfy the Hamiltonian equation with Eq.~(\ref{eqn:TLHam}). Therefore, following the canonical quantization of a harmonic oscillator, the current and voltages on the line can be given in terms of the bosonic ladder operators. Specifically, the current at the position of the qubit is
\begin{align}
    I_s(z=q,t)&=\sqrt{\frac{\hbar\omega}{2l_0d(1+|r|^2)}}\l\{ia^\dagger\l[\exp{(-i\omega_sz_q/v)}\r.\r.\nonumber\\
    &\l.\l.-r_s\exp{(i\omega_sz_q/v)}\r]+\t{h.c.}\r\}.
\end{align}

The coupling strength between the qubit and a mode on the open TL is proportional to the product of the quantum fluctuation of the current at the position of the qubit and the junction phase matrix element, given by
\begin{align}
    g_s =\phi_{0}\left|\gamma_{5, 10}\right|\sqrt{\frac{\hbar\omega_s}{2l_0d(1+|r_s|^2)}}\nonumber\\
    \times|\exp{(-i\omega_sz_q/v)}-r_s\exp{(i\omega_sz_q/v)}|.\label{eqn:GsFilter}
\end{align}
Finally, the relaxation rate can be obtained by inserting Eq.~(\ref{eqn:GsFilter}) into Eq.~(\ref{eqn:SBGamma}), and take the $d\xrightarrow[]{} \infty$ limit, which gives 
\begin{align}
    \Gamma(\omega)&=\frac{2\pi}{\hbar^2}\sum_{n,s}g_{n,s}^2\delta(\omega-\omega_{n,s})\\
    &=\sum_s\frac{\phi_{0}^2\left|\gamma_{5, 10}\right|^2\omega}{2\hbar Z_0(1+|r_s|^2)}|\exp{(-i\omega z_q/v)}\nonumber\\
    &-r_s\exp{(i\omega z_q/v)}|^2.\label{eq:FilterRelaxation}
\end{align}

To make the estimate concrete, we consider the component $F$, placed at a distance $z_q=20~\t{cm}$ from the qubit, to be reciprocal and characterized by a voltage standing wave ratio (VSWR), which allow us to obtain the ratio between left and right travelling wave amplitudes. We can calculate the radiative relaxation rate $\Gamma_1$ for several values of VSWR to observe the effects of finite reflections in the component $F$ on the coupling in the system. As shown in the Sec.~\ref{sec:discrepancy} of the main text, very large reflections would be required to explain the disagreement observed between the calculated and measured coupling values.

\end{document}